# Surface Reconstructions of Heusler Compounds in the Ni-Ti-Sn (001) System

A.D. Rice[1], A. Sharan[2], N.S. Wilson[1], S.D. Harrington[1], M. Pendharkar[4], A. Janotti[2,3] and C.J. Palmstrøm[*,1,4]

[1]Materials Department, University of California, Santa Barbara, California 93106, USA

[2]Department of Physics & Astronomy, University of Delaware, Newark, Delaware 19716, USA

[3]Department of Materials Science & Engineering, University of Delaware, Newark, Delaware 19716, USA

[4]Department of Electrical & Computer Engineering, University of California, Santa Barbara, California 93106, USA

**Abstract**

As progress is made on thin-film synthesis of Heusler compounds, a more complete understanding of the surface will be required to control their properties, especially as functional heterostructures are explored. Here, the surface reconstructions of semiconducting half-Heusler NiTiSn(001), and $Ni_{1+x}TiSn(001)$ (x=0.0-1.0) are explored as a way to optimize growth conditions during molecular beam epitaxy. Density functional theory (DFT) calculations were carried out to guide the interpretation of the experimental results. For NiTiSn(001) a c(2x2) surface reconstruction was observed for Sn rich samples, while a (1x1) unreconstructed surface was observed for Ti-rich samples. A narrow range around 1:1:1 stoichiometry exhibited a (2x1) surface reconstruction. Electrical transport is used to relate the observed reflection high energy electron diffraction (RHEED) pattern during and after growth with carrier concentration and stoichiometry. Scanning tunneling microscopy and RHEED were used to examine surface reconstructions, the results of which are in good agreement with density functional calculations. X-ray photoelectron spectroscopy was used to determine surface termination and stoichiometry. Atomic surface models are proposed, which suggest Sn-dimers form in reconstructed $Ni_{1+x}TiSn(001)$ half-Heusler surfaces (x<0.25) with a transition to Ni terminated surfaces for x > 0.25.

## 1. Introduction

Heusler compounds, ternary compounds which span most of the periodic table, have been predicted to be semiconducting[1], metallic, magnetic, half-metallic,[2] superconducting,[3] and some to contain topological surface states[4] depending on their composition. Additionally, many of these compounds can be grown as thin films, have lattice constants near common III-V substrates and contain elements which can be thermally evaporated from effusion cells.[5] This, combined with their wide range of properties, suggests the possibility of creating multifunctional heterostructures by combining Heusler compounds with their similar crystal structures and symmetries; however, their properties depend critically on stoichiometry.[6,7,8] The stoichiometry issue has been overcome for some materials systems by utilizing a "growth window" in which the excess amount of a component with a high vapor pressure does not incorporate during molecular beam epitaxy (MBE), such as for the growth of GaAs[9] and $PbTiO_3$.[10] In general, however, Heusler compounds contain two or more low vapor-pressure elements with sticking coefficients close to unity at any reasonable growth temperature. Hence, exact stoichiometry can only be obtained by controlling the individual elemental fluxes very carefully. If the surface reconstruction depends on stoichiometry, it may be possible to use the surface reconstruction and any changes in it to determine and control the relative elemental fluxes to achieve the desired bulk stoichiometry. *In-situ* reflection high-energy electron diffraction (RHEED) during MBE growth allows for observation of the surface reconstruction, providing real-time feedback.

---

[*] Corresponding email address: cpalmstrom@ece.ucsb.edu

Previous work on CoTiSb[11] and PtLuSb[12] surfaces has taken advantage of the relatively high vapor pressure of Sb to change the surface reconstruction by varying the substrate temperature and Sb overpressure without varying the bulk stoichiometry, analogous to conventional III-V semiconductors.[13,14] While the half-Heusler (hH) NiTiSn, a semiconductor with a predicted band gap of around 0.45 eV,[15] contains no similarly high vapor pressure elements, stoichiometry changes may still be reflected in the surface reconstruction. Previously, epitaxial $Ni_{1+x}TiSn$ films have been grown by MBE on MgO(001) substrates.[16] Films with $x \leq 0.25$ ($Ni_{1.25}TiSn$) showed (2×1)/(1×2) RHEED patterns similar to films at 1:1:1 stoichiometry. The (2x1)/(1x2) observed surface reconstruction resulted from a mixture of 90° rotated (2x1) domains due to the 4-fold surface symmetry of the MgO(001) surface. However, for films with $x > 0.25$ the ½ order streaks disappeared with the electron beam incident in the <110> directions. As the Ni excess increased, the 1st order streaks also began to disappear in this direction, while the RHEED patterns with the electron beam in the <100> directions appeared unchanged. This is indicative of a change in the surface reconstruction unit cell resulting in a smaller 45° degree rotated surface unit cell.[16]

In this work, the relationship between surface reconstructions of NiTiSn(001) films and their stoichiometry with emphasis on changes in the Ti/Sn ratio is presented. *In-situ* X-ray photoelectron spectroscopy (XPS) and scanning tunneling microscopy (STM) are used to study the chemical and structural properties of these surfaces, leading to proposed models of these surfaces. DFT calculations are performed to understand the changes in surface reconstruction as a function of surface stoichiometry; the resulting stable surface reconstructions are used to match the STM results. Results on the full-Heusler (fH), $Ni_2TiSn$, and $Ni_{1+x}TiSn$ films are also shown and examined further to present a complete picture of surface termination in the Ni-Ti-Sn(001) system.

## 2. Experimental

NiTiSn films were grown using MBE in a VG-V80 MBE growth chamber with a base pressure of $<1\times10^{-10}$ Torr as previously reported.[16,17] Elemental sources were evaporated from effusion cells at a growth rate of $3\times10^{16}$ atoms/cm²-hr for each element, corresponding to a growth rate of around 15 nm/hr, on 10 mm × 10 mm × 0.5 mm MgO(001) substrates from Crystec GmBh. The effusion cell sources were calibrated using Rutherford backscattering spectrometry (RBS) to determine the relationship between actual elemental flux and an ion gauge measurement of the beam equivalent pressure (BEP). This was done by measuring the deposited number of atoms/cm² for each element with a number of different cell temperatures for calibration samples grown on Si substrates. For non-stoichiometric films, the Sn flux was held constant, with the Ti BEP being raised ~1% per sample (corresponding to the cell temperature adjusted by 1°C) and the Ni cell flux adjusted by changing its cell temperature to maintain the Ni:(Ti+Sn) ratio of 1:2 (i.e. $Ni_{1+y}Ti_{1+2y}Sn$, which is equivalent to $NiTi_{1+\delta}Sn_{1-\delta}$ with $\delta=y/(1+y)$). $Ni_{1+x}TiSn$ samples were also grown using a similar process, with the Ni cell flux set as needed while keeping the Ti and Sn fluxes constant. Samples were grown at a substrate temperature of approximately 370 °C as determined by the arsenic desorption temperature of As-capped $In_{0.52}Al_{0.48}As/InP(001)$ test structures. All samples were post-growth annealed at approximately 450 °C for 15 minutes. Films for transport measurements were then capped with 5 nm AlO$_x$, deposited by e-beam evaporation of Al$_2$O$_3$ at room temperature. The carrier concentration was measured on samples using the Hall effect with annealed indium contacts in the Van der Pauw geometry with magnetic fields up to 5kOe.

For *in-situ* STM studies, the substrates used were highly conductive n-type doped (>1x10$^{18}$ Si/cm$^3$) In$_{0.52}$Al$_{0.48}$As/In$_{0.53}$Ga$_{0.47}$As/In$_{0.52}$Al$_{0.48}$As (50/300/50 nm) structure on n-type InP(001) with an additional 5 nm thick epitaxial GdAs layer all grown by MBE. This GdAs layer acts as a diffusion barrier (similar to the ErAs system)[18] between the NiTiSn and the III-V semiconductor layers. Nickel has been shown to be reactive with III-V semiconductors at viable growth temperatures [19]. The III-V semiconductor structures with the GdAs diffusion barrier were grown in a separate VG V80H III-V system and capped with amorphous arsenic. Samples were then cleaved *ex situ* and prior to each NiTiSn epitaxial layer sample growth, mounted on STM sample holders using indium and loaded into the interconnected ultra-high vacuum (UHV) growth and characterization system with base pressures <1x10$^{-10}$ mbar throughout. The arsenic cap was then thermally desorbed in a transfer chamber immediately prior to growth. The STM measurements of the NiTiSn layers were performed *in-situ* using an Omicron LT-STM at 77 K.

XPS was also performed *in situ* by using an interconnected UHV system. An Al Kα source (1486.7 eV) was used to scan a region in binding energy from 450 eV to 900 eV to obtain Ni, Ti, and Sn core level peaks. The XPS spectra were fitted using a convolution of Gaussian and Lorentzian line shapes. To model intensities, the methods outlined by Schultz et al [18] were used, which treats the signal as the sum of N attenuated atomic layer contributions,

$$I_x^{tot} = \sum_{n=1}^{N} I_x^i \qquad (1)$$

$$I_x^i = f_x \sigma_x n_x^i \prod_{i=1}^{N} \alpha_x^i \qquad (2)$$

$$\alpha_x^i = e^{\frac{-d_i}{\cos(\theta) * \lambda_{x,i}}} \qquad (3)$$

where I$_x$ is the intensity of elemental peak, x, N is the total number of layers, f$_x$ is instrumental factors, σ$_x$ is the cross section, n$^i_x$ is the atomic density, d$_i$ is the atomic layer spacing, θ is the angle of incidence (55° in this experiment), and λ$_{x,i}$ is the electron mean free path through layer i. Each of these factors are specific to element x and must be calculated separately. To minimize instrumental factors, analysis was done on peak area ratios normalized by the same ratios with the spectra from a NiTiSn surface with a (1x1) surface reconstruction. Electron mean free paths, λ, were calculated using inelastic mean free paths, λ$_{IMFP}$, from the Tanuma, Powell, and Penn [20] and adding the following elastic correction:[21]

$$\lambda = \lambda_{IMFP}(1 - 0.028Z^{0.5})[0.501 + 0.068 \ln(E)] \qquad (4)$$

Surface coverages from proposed ball and stick models were used in the model calculations to obtain the relative elemental XPS peak ratios. Changes in Ti and Ni bulk composition were also accounted for in the calculations as needed.

The DFT [22,23,24] calculations of surface reconstructions are based on the generalized gradient approximation (GGA) functional of Perdew, Burke and Ernzerhof (PBE)[25] as implemented in the VASP code. [26,27] The interactions between the valence electrons and ionic cores are treated using the projector-augmented wave potentials. [28,29] The electronic structure of bulk NiTiSn is in good agreement with previous studies, [30] with a calculated band gap of 0.45 eV. The surface calculations are performed using a 4x4 slab geometry with 13 atomic layers, with two equivalent top and bottom surfaces, rotated 90° with respect to each other due to symmetry of the underlying NiSn zincblende sublattice, and a vacuum region of 12 Å thick. The integrations over the Brillouin zone were performed using a 2x2x1 mesh of special *k*-points, and the energy cut off for plane wave expansion was set to 270 eV. The atoms on the top and bottom four atomic layers were allowed to relax, while the atoms in the middle layers of the slab were fixed to their ideal bulk positions. We focused on the TiSn terminated layer, so the ideal surface contains 16 Ti and 16 Sn atoms. The surface stoichiometry was varied by changing the concentration of Ti atoms on the surface. The stoichiometry in the bulk remained 1:1:1 for Ni, Ti, and Sn atoms. Simulated STM images of the surface reconstructions were visualized using the Hive-STM package [31].

## 3. Results

### 3.1 Half-Heusler Surfaces

Figure 1(a) shows RHEED patterns for NiTi$_{1+\delta}$Sn$_{1-\delta}$(001) (-0.05≤δ≤0.02) films taken with the electron beam incident along the <100> and <110> directions at 250° C following a post-growth anneal for samples of varying values of δ. A (2×1)/(1×2) surface reconstruction (with the surface unit cell defined with axes along <110>, as is typical for III-V(001) surfaces), which will be referred to as a (2×1) reconstruction, was observed for stoichiometric NiTiSn in agreement with previous results.[17] For Ti excess of >1% (δ > 0.01), the ½ order streaks with the electron beam along <110> disappear and an unreconstructed (1×1) surface reconstruction was observed. For larger Sn excess (δ = -0.03– -0.05), ½ order streaks were observed only in the <100> type directions, corresponding to a c(2×2) surface reconstruction. For small amounts of Sn excess (δ = -0.01– -0.02), ½ order streaks were observed in both <100> and <110> type directions and will be referred to as a (2×2) surface reconstruction. During these growths, streaks were only visible in either set of directions, with the additional streaks appearing only after the post-growth anneal in the other set of directions. Figure 1(b) summarizes these patterns in reciprocal space. Figure 2 shows the approximate surface reconstruction phase diagram as a function of Ti composition and substrate temperature. A c(2×4) reconstruction was also observed at low temperatures and slight Ti excess.

To determine systematic effects of off-stoichiometry and to compare RHEED patterns with electronic properties, room-temperature Hall measurements were performed to determine the carrier concentration. Figure 3 shows the carrier concentration as a function of composition. The relative compositions cited throughout are believed to be accurate to 1%. A minimum in the carrier concentration is found for stoichiometric NiTiSn films. For excess Ti, the measured carrier concentration increases sharply, while for Sn excess, the increase appears more gradual.

Figure 4(a) shows XPS scans of the energy region of interest for each of the previously observed half Heusler reconstructions. The Ni 2p, Ti 2p, and Sn 3p peak areas were used to determine relative differences in the surface stoichiometry. Figure 4(b) shows the measured peak area ratios of Ti 2p/Ni 2p and Sn 3p/Ni 2p as a function of Ti composition (1+δ) in NiTi$_{1+\delta}$Sn$_{1-\delta}$ from these scans. These ratios are normalized by the values for the (1×1) sample. Error bars are estimated from the statistical uncertainty and the uncertainty in the peak area fits, which propagate through as the ratios are taken. Ni/Sn ratio is largely unchanged, with only a slight increase in the c(2×2) and (2×1) reconstructions, while the Ni/Ti ratio increases to 1.20 for the c(2×2) reconstructions and 1.52 for the (2×1) reconstruction.

Figures 5(a), 5(b), and 5(c) show filled-states STM images taken on terraces having the c(2×2) (NiTi$_{0.97}$Sn$_{1.03}$), (2×1) (NiTiSn), and c(2×4) (NiTi$_{1.01}$Sn$_{0.99}$) surface reconstructions respectively. Fig 5(d), 5(e) and (f) shows the corresponding simulated STM images using DFT. The (2×1) and c(2×4) surface also showed domains rotated by 90°. The c(2×2) surface shows a clear checker board pattern in the STM image (Fig. 5(a), 5(d)), which is consistent with the c(2×2) surface reconstruction, the (2×1) shows rows running along the <110> directions (Fig 5(b) and Fig 5(e)). The (2×1) surface also showed regions appearing to transition from a (2×2) to a c(2×4) surface. The reconstructions were measured to have the same 2× spacing of approximately 0.84 nm (twice the surface unit cell side length, equal to $\frac{a_{hH}}{\sqrt{2}} * 2$). As can be seen in Fig. 5, there is a good agreement between simulated and measured STM images. Figure 5 (g-i) shows height contour line profiles characteristic of the labeled directions in the STM image. Steps between terraces throughout the sample were intervals of around 0.3 nm, in good agreement with bilayer atomic steps, $\frac{a_{hH}}{2}$ (=0.296 nm). The (2×1) reconstruction contained 40 pm corrugations between rows, while the c(2×2) also had 20 pm corrugations going diagonally along the Ni rows. The c(2×4) surface showed much sharper features, with larger measured depths between dimers (Fig. 5(i)).

### 3.2 Full-Heusler Surfaces

Figure 6(a) shows RHEED patterns from a full Heusler Ni$_2$TiSn surface. In comparison with the (1×1) RHEED pattern in Fig. 1 for NiTi$_{1+\delta}$Sn$_{1-\delta}$, it is clear the RHEED patterns with the electron beam incident along <100> are similar, but incident along <110> are different, with the 1$^{st}$ order streak in Fig. 1 becoming a ½ order streak in Fig. 6 while also having a reduced intensity. This difference can be explained by two different surface unit cells rotated by 45° with respect to each other, with the one for Ni$_2$TiSn being 1/√2 smaller than for a Ti-Sn terminated surface layer. Note that the hH (1x1) unit cell would correspond to a c(2x2) unit cell using this smaller surface unit cell definition, Fig. 6(c). This is consistent with a Ni-terminated or a disordered Ti-Sn terminated full Heusler surface.[16]

Figure 7 shows the Ni/Ti and Sn/Ti peak area ratios for Ni$_{1+x}$TiSn films as a function of the bulk film Ni composition, normalized by (1×1) surface reconstruction values. Little change is observed in the Sn/Ti ratio, while the Ni/Ti ratio for fH-like films increases gradually with Ni excess. There is a sharp increase in the Ni/Ti ratio for NiTiSn (001) (x=0 and δ=0) with a (2×1) surface reconstruction. The dashed lines in Fig. 7 correspond to values predicted based on assumptions that an abrupt transition occurred above $x_{Ni}$=1.25 from a hH-like (2x1) to a fH-like surface reconstruction, as observed in RHEED. The two dashed blue lines above $x_{Ni}$=1.25 correspond to Ni/Ti ratios from models either assuming a Ni terminated surface or a Ti-Sn terminated surface of a Heusler film, both with increasingly filled Ni sites (in the case of Ni-termination, this top Ni layer is also assumed to be partially filled proportional to the amount of excess Ni). The dotted orange line corresponds to Ni/Ti ratios assuming a Ni terminated fH, which has double the Ni atomic density per atomic layer. Both models account for the additional Ni throughout the bulk film, while the Ni model also adds an additional x/2$_{Ni}$ monolayers (relative to full-Heusler occupation of these sites) of Ni termination to the surface. These models will be discussed in more detail in the discussion section below.

Figure 8 shows a filled state STM image of a fH surface. This Ni$_2$TiSn surface has rows running in the <010> directions with a measured row spacing consistent with a smaller, rotated (2×1). These fH reconstruction rows show 10-15 pm deep trenches between rows, much smaller than that measured on the hH surfaces. Other terraces on the surface also showed similar rows but rotated by 90°.

## 4. Discussion

The RHEED patterns suggest that the observed surface reconstruction depends strongly on composition (Fig. 1). For even small amounts of Ti excess (δ > 0.01) in NiTi$_{1+δ}$Sn$_{1-δ}$(001), the ½ order <110> streaks disappeared in the RHEED images, resulting in an unreconstructed (1×1) surface unit cell, while a (2×2) reconstruction was observed for the case with a Ti deficiency of δ = -0.02 and a c(2×2) for δ = -0.05. Some "splitting" of the RHEED streaks in <100> directions was observed for the Ti deficient (δ <0) samples, suggesting some roughening with the formation of corrugations on the surface.[32] The amount of this texture increased with the Ti deficiency (Sn excess, δ<0) and was not observed for stoichiometric or Ti-rich (δ>0) samples. The observed (2×2) surface reconstruction for δ ~ -0.02 is most likely a mixture of c(2×2) and (2×1)/(1×2) reconstructions, as seen in Fig. 1(b). This particular RHEED pattern was only observed post-growth after annealing and cooling the samples to room temperature. During growth these samples showed either a c(2×2) or a (2×1)/(1×2) reconstruction. The (2×1) surface can also be seen to have a faint set of diagonal lines converging at each ½ order point in the <100> direction, which emerged upon annealing and subsequent cooling. This suggests a transition occurs at lower temperatures, with some amount of roughening also accompanying it.

Room temperature Hall measurements showed minimum carrier density near stoichiometry, with strong dependence on samples with Ti excess (δ>0) and a small dependence for samples with Ti deficiency (Sn excess, δ<0). This may be explained by the ternary phase diagram, which shows a phase field with low mutual solubility between elemental Sn and NiTiSn[33]. Hence, if the excess Sn is expected to phase separate rather than incorporating as an electrically active donor state, this could have a reduced effect at low concentrations on the measured carrier concentration. The samples with the sharpest (2×1)/(1×2) RHEED pattern corresponded to the ones with lowest observed electron concentration, and stoichiometric δ=0 as determined from the RBS calibrations measurements.

To be able to propose atomic models for the different surface reconstructions, the atomic composition of the surface is needed. Figure 4(b) shows the Ni 2p/Ti 2p and Ni 2p/Sn 3p normalized peak area ratios for NiTi$_{1+δ}$Sn$_{1-δ}$ films with -0.04≤δ≤0.02. The normalization is made using a sample with the (1×1) surface reconstruction, which is assumed to correspond to a bulk Ti-Sn terminated hH surface. Since the Ni 2p/Ti 2p ratio is more sensitive to δ than the Ni 2p/Sn 3p, it is suggestive that surface reconstructions involve additional adlayers of Ni and Sn on top of a Ti-Sn bulk terminated surface. The XPS data gives information regarding the surface stoichiometry by showing trends in elemental composition and confirming basic assumptions regarding surface termination. The intensity for a specific elemental peak, x, in the XPS spectrum is the solution to an infinite series summation due to the film being much thicker than the photoelectron mean free path, λ:

$$I_x^{tot} = \sum_{n=1}^{\infty} I_x^i = f_x \left[ \sigma_x n_x^{TiSn} + \sigma_x n_x^{TiSn} \alpha_x^{TiSn} \right] * \frac{1}{1-\alpha_x^{TiSn}\alpha_x^{Ni}} \qquad (5)$$

Surface reconstructions add attenuation factors, $(α_x)^m$, where m is the number of monolayers, as well as additional intensity terms using eq. (2). The asterisks in Fig. 4(b) are calculated values from these equations assuming a bulk Ti-Sn terminated unit cell, followed by a monolayer of Ni, followed by one additional monolayer Sn on the (2x1)/(1x2) surface, analogous to CoTiSb.[11] This is done by using eq. (5) for the "bulk" signal and adding additional terms in the form of eq. (2), as well as attenuation factors to eq. (5). An additional ½ monolayer of Ti is placed within the top surface layer of the c(2x2) surface (these values come from ball and stick models to be discussed later). These values largely match up well with the observed values and are consistent with the formation of Sn dimers above a layer of Ni. Some XPS peak area ratios change for different reconstructions despite little change in relative composition due to differences in the attenuation term in eq. (5) arising from different kinetic energies (and thus mean free paths).

It is informative to compare the NiTiSn(001) with CoTiSb(001). In the case of CoTiSb it has been suggested that it may be thought of as zincblende $CoSb^{4-}$ with $Ti^{4+}$, which would lead to $Co^{1-}$ and $Sb^{3-}$.[34] This leads to electron counting model for stabilization of different surface reconstructions with partial Ti coverage and Sb dimer formation[11]. Following this argument, NiTiSn would correspond $NiSn^{4-}$ stuffed with $Ti^{4+}$, resulting in $Ni^0$, $Sn^{4-}$ and $Ti^{4+}$ and the surface reconstruction would result from Sn-dimer formation. The (2x1) cell would contain one Sn-Sn dimer bond, two Sn dangling bonds, and four Ni-Sn back bonds. This tetrahedral bonding picture for the p-d hybridized NiSn sublattice is consistent with real space electronic structure calculations.[34] Filling up these bonds requires a total of 2 (Sn dimer) + 8 (Ni-Sn back) + 4 (Sn dangling) + 10 (Ni d10) = 24 electrons per (2×1) cell, assuming that each Sn dangling bond is fully occupied (2 electrons each). For half-filled dangling bonds (1 electron each), 22 electrons would be required. The number of electrons available is 2×4 from Sn in the top layer, 2×4×$n_{Ti}$ from Ti in the top layer ($n_{Ti}$ is the fractional occupancy of Ti at the surface), and $2 \times \frac{10}{2}$ from Ni in the second layer, where the divide by two is to avoid double counting the Ni atoms. This results in a total of $18 + 8n_{Ti}$ electrons available. Filling the bonds require 2 electrons for the dimer bond, 8 for the Ni-Sn bonds, and 10 for the Ni atoms, for a total of 20 electrons, without accounting for the 2 dangling Sn orbitals. These conditions would be fulfilled for ¾ and ½ Ti occupancy for filled and half-filled dangling bonds, respectively. For CoTiSb the corresponding Ti coverage would be 5/8 and 3/8 for the filled and half-filled dangling bonds[11], respectively. This arises due the difference in the valence between Sb and Sn and Co and Ni[11].

The STM images confirm the RHEED results, with Sn-rich films and stoichiometric films showing STM images corresponding to the surface unit cells. The image of the (2×1) surface helps address the issue of the (2×2) transition addressed earlier. The majority of the surface clearly shows a zig-zag structure that would increase the periodicity of the unit cell relative to the highlighted unit cell. It should be noted, however, that the RHEED patterns were only observed down to 200 °C, while STM images were taken at 4 K, and DFT is performed at 0 K. As suggested earlier, even at stoichiometry, a transition from the (2×1) to (2×2) surfaces may still be occurring, but at temperatures lower than observed for the more Sn-rich surfaces. The roughness and disorder of the dimers is likely indicative of an incomplete transition, which may not be possible to complete at higher temperature ramp rates.

By combining the STM images with the XPS results, electron counting and DFT calculations, we propose atomic models for different NiTiSn (001) surface reconstructions (Fig. 9). Figure 9(a) shows a model for the c(2x2), filling in half of the missing Ti surface sites compared to the (2×1) surface, and with dimers alternating rather than forming rows. On top of a bulk terminated Ti-Sn layer there is a layer of Ni followed by a layer of Sn dimers with half of the Ti sites vacant, which would be consistent with half-filled Sn-dimer dangling orbitals and the higher Ni 2p/Ti 2p ratio in comparison to the (1×1) surface. It is also consistent with the steps observed in the STM images, with larger vertical dips in the <110> direction, appearing to correspond to ½ unit cell (2 atomic layers) deep trenches compared to <100> directions, which would be consistent with ¼ unit cell (1 atomic layer) deep trenches. The model in Figure 9(b) for the (2×2) surface shows a Sn-dimer structure along columns with an in-plane zigzag structure. There is also sizeable out of plane buckling, ~0.6 Å, of the Sn-Sn dimers on the surface. The top surface is made up of a bulk terminated Ti-Sn layer with a Ni layer, and then a top surface layer consisting of Sn dimers with all of the Ti sites vacant. A (2×1) model would look similar, with reduced buckling in both directions, producing straight dimers that reduce the periodicity of the surface.

The c(2x4) reconstruction appears to be stable at scanning temperatures, with very sharp features. Given the partial coverage of the surface, no XPS data for the surface is available, which limits the ability to determine Ti content in the surface layer. Figure 9(c) shows a model with 2/8 Ti sites per unit cell filled in the surface layer. While no chemical sensitive technique has been used due to the partial coverage of this surface, this model is consistent with the STM images and electron counting. For the STM image of the c(2×2) surface, large mounds (200+pm, circled in Fig. 5(a)) are seen. These are potentially due to accumulation and agglomeration of excess Sn on the surface. These features would also not have sufficient surface coverage to be detected by XPS. No significant out of plane buckling of Sn dimers has been observed for c(2x2) and c(2x4) surface reconstructions in the DFT calculations.

The fH surface, shown in Fig. 8(a), showed rows rather than a square unit cell as previously suggested. While this observed surface reconstruction would only remove 1st order streaks in the [010] direction, other terraces on this surface also showed these rows rotated 90°, which would only remove 1st order streaks in the [100] direction. The small intensity still present in these streaks is likely the result of the underlying Ti-Sn layer. Height profiles between rows showed much smaller values than between the other dimer rows, suggesting that these perturbations may be electronic in nature, as STM is sensitive to both height and local electron density of states. This combination of domains would still be consistent with a Ni-terminated surface, however with a slightly different structure than previously proposed[16].

XPS data of Ni-rich samples showed a trend consistent with Ni termination, which, compared to the (1×1) surface, would expect very little change in the Sn/Ti ratios and large increases in the Ni/Ti ratio. The lower of the dotted blue lines in Fig. 8, corresponding to a Ti-Sn termination, greatly under predicts the Ni signal. The Ni model (upper blue line) does very well for each pure surface, though slightly over-predicts the Ni/Ti ratio in the middle region, in particular not showing a discontinuity during the transition from a Ni-rich (2×1) at $x_{Ni}$=1.25 that is assumed to happen abruptly from the observed RHEED patterns. While the Sn/Ti ratios are in good agreement with calculations throughout, the values are very similar to the (1x1) surface. This would make it hard to use this ratio to distinguish between a purely Ni-terminated surface (like the model assumes) and a film surface also containing areas of Sn-Ti termination, which would decrease the Ni/Ti ratio relative to the model's prediction. Despite this, the XPS data provides further evidence that the fH reconstruction shows Ni termination. This change in termination may likely be present in other fH systems with similar RHEED patterns, such as Co2TiGe[35] and Co2MnSi[36] as well as other related systems[37].

To summarize these data, Fig. 9 shows ball and stick models for these discussed surfaces. In Fig. 9(a-c), the Sn-excess is reflected by a Ti-deficient surface in which Sn dimers are formed. The Ti vacancy sites referenced earlier are highlighted by dotted circles (sitting above Sn atoms in a lower layer). These models are generally similar, with a termination in the Ti-Sn plane and Sn dimers. Changes reflect observed differences in dimer structures and Ti content in STM and XPS data. While in the case of 9(b), the unit cell as drawn would produce a (2×2) reconstruction, however in practice, the existence of disorder in the buckling of the dimers would produce observed (2×1). For fH-like surfaces, a (2×1) model reflecting the rows seen in STM is shown in Fig. 9(d). Given the small height differences between rows, and the metallic nature of the phase, only slight displacements in the Ni atoms towards each other to break the symmetry are shown rather than dimer formation, and, as stated earlier, these features may be ultimately electronic in nature rather than structural given their magnitude.

## 5. Conclusions

Using RHEED, STM, XPS data and DFT calculations, surface reconstruction models are proposed for the Ni-Ti-Sn Heusler system. These appear to depend on whether it is a Sn, Ni, or SnTi terminated surface, and on the amount of Ti-vacancies for surfaces with Sn dimers. STM and XPS data also confirms the previously proposed surface termination for $Ni_2TiSn$, although it appears to be a (2×1)/(1×2) reconstruction rather than a (1x1), based on the unit cell for nickel termination. As thin-film growth of Heusler compounds becomes more sophisticated, this may have implications on device performance as heterostructures are explored. Regardless, these reconstructions give relatively sensitive feedback regarding the Ti:Sn ratio in the films, providing immediate information regarding growth quality in absence of an obtainable growth window.

**Acknowledgements**

The preliminary epitaxial growth studies were supported by the MRSEC Program of the National Science Foundation (DMR-1121053) with the later epitaxial growth and theory studies by the U.S. Department of Energy Basic Energy Science program (DE-SC0014388), the STM by the National Science Foundation (DMR-1507875), the XPS studies by ONR Vannevar Bush Faculty Fellowship (NSSEFF) (ONR N00014-15-1-2845). S.D.H. was supported in part by the National Science Foundation Graduate Student Fellowship under Grant No. 1144085. A.S. and A.J. acknowledge the use of resources of the National Energy Research Scientific Computing Center (NERSC), a U.S. Department of Energy Office of Science User Facility operated under Contract No. DE-AC02-05CH11231. We also acknowledge the use of facilities within the LeRoy Eyring Center for Solid State Science at Arizona State University.

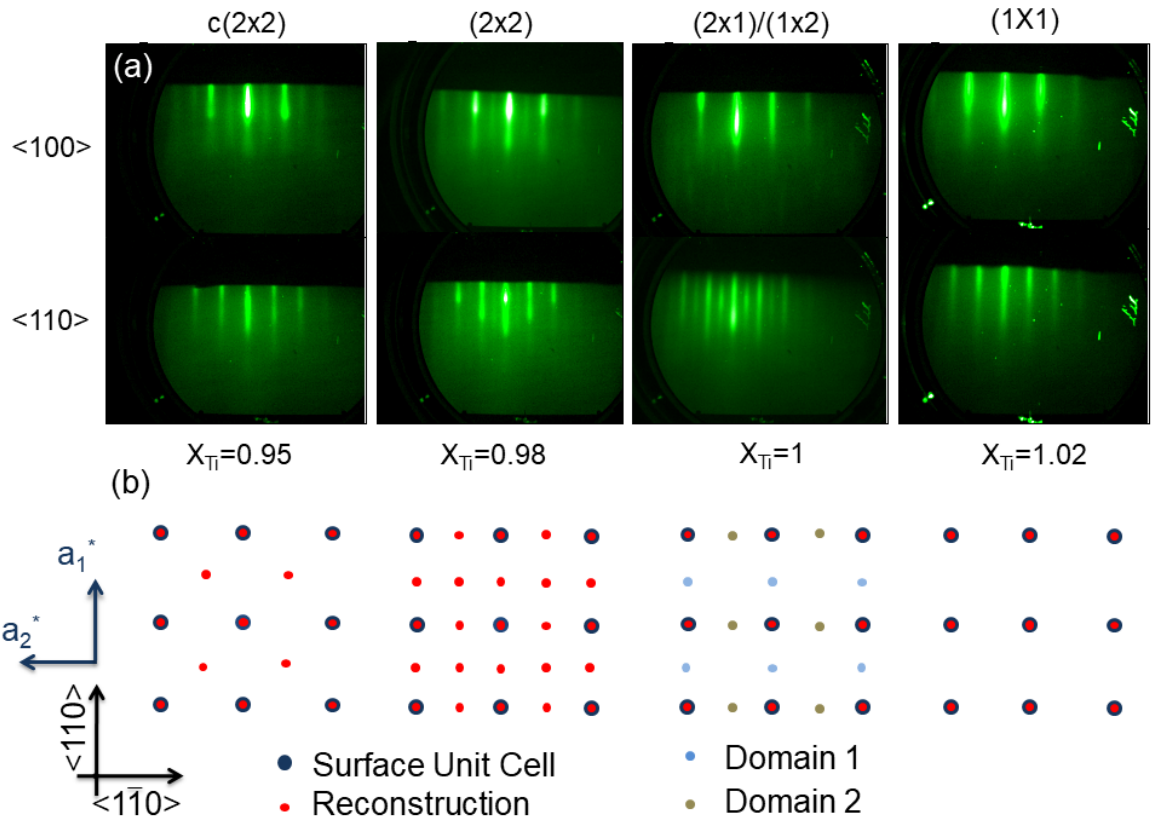

Figure 1. (a) RHEED images of 20 nm thick NiTi$_{1+\delta}$ Sn$_{1-\delta}$ with varying δ (X$_{Ti}$ = 1+ δ ) and (b) reciprocal space of surfaces (the large dark blue dots correspond to the reciprocal lattice of the bulk terminated (1x1) surface unit cell and the smaller dots to the additional reciprocal lattice rods from the reconstructed surface unit cell. Due to the 4-fold symmetry of the substrates, the (2x1) surface is shown to have 2 different orientations at 90° to each other. a$_1^*$, a$_2^*$ correspond to basis vectors for the surface unit cell.

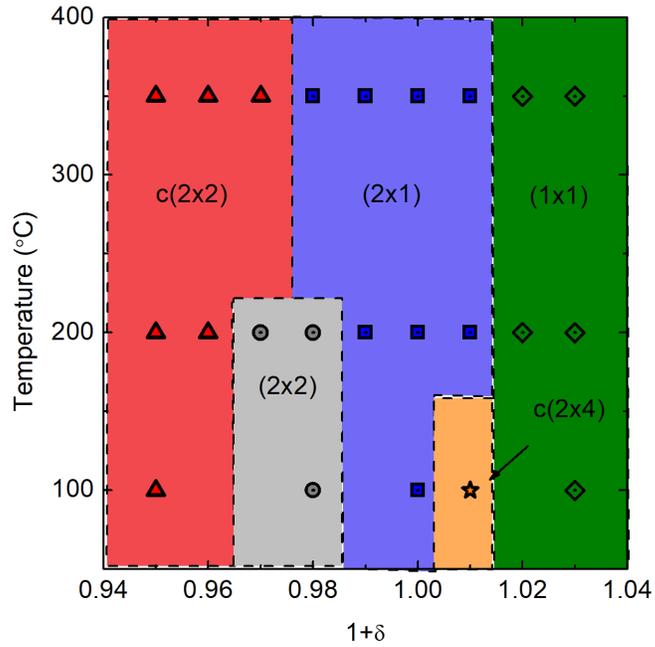

Figure 2. Approximate surface reconstruction phase diagram as a function of temperature and bulk Ti composition (NiTi$_{1+\delta}$Sn$_{1-\delta}$). Points represent experimental observations.

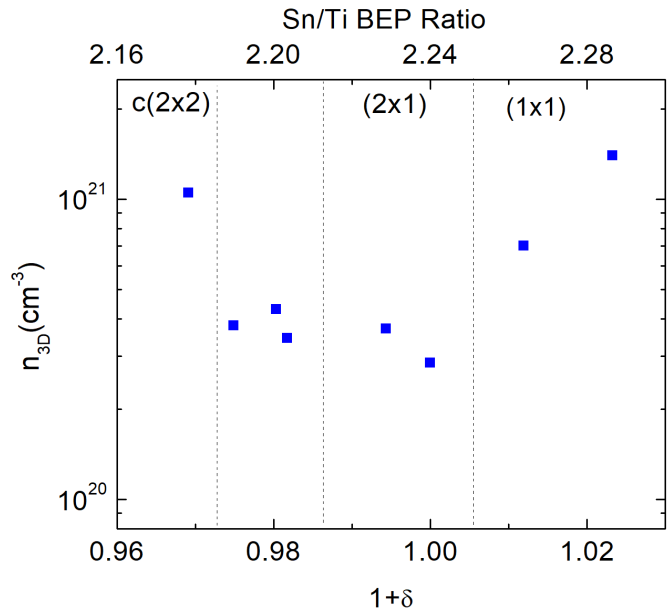

Figure 3. 3D carrier concentration as measured by the Hall effect for NiTi$_{1+\delta}$Sn$_{1-\delta}$ epitaxial films. BEP ratio is shown on the top, which is then converted to composition (bottom axis) using RBS calibration.

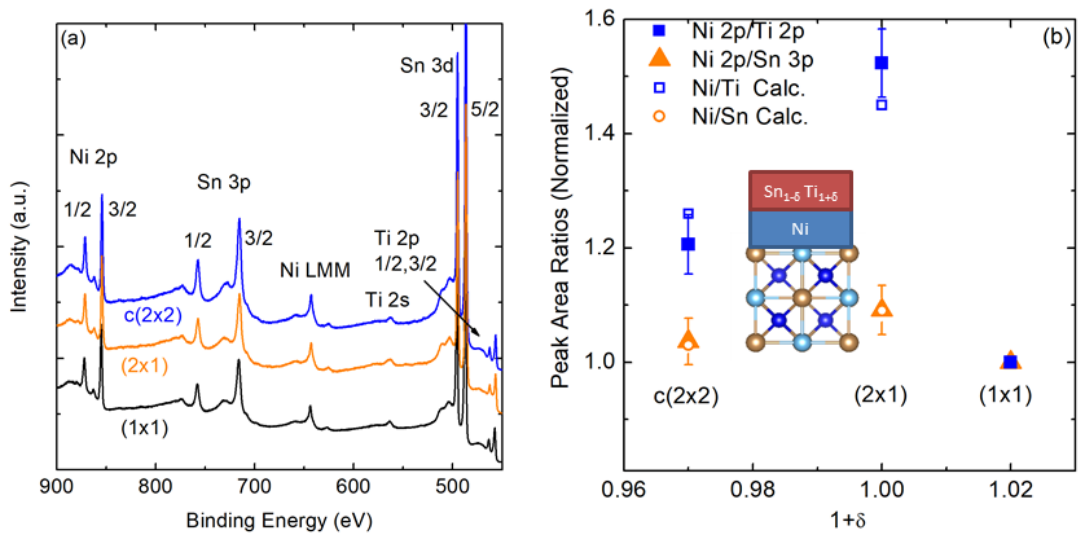

Figure 4. (a) XPS scans of NiTi$_{1+\delta}$ Sn$_{1-\delta}$ films taken for various surface reconstructions and (b) resulting peak area ratios. Ratios show two elemental peak areas relative to each other, and then normalized to the same ratio for the (1x1) surface. Unfilled symbols correspond to calculated normalized ratios from proposed surface models.

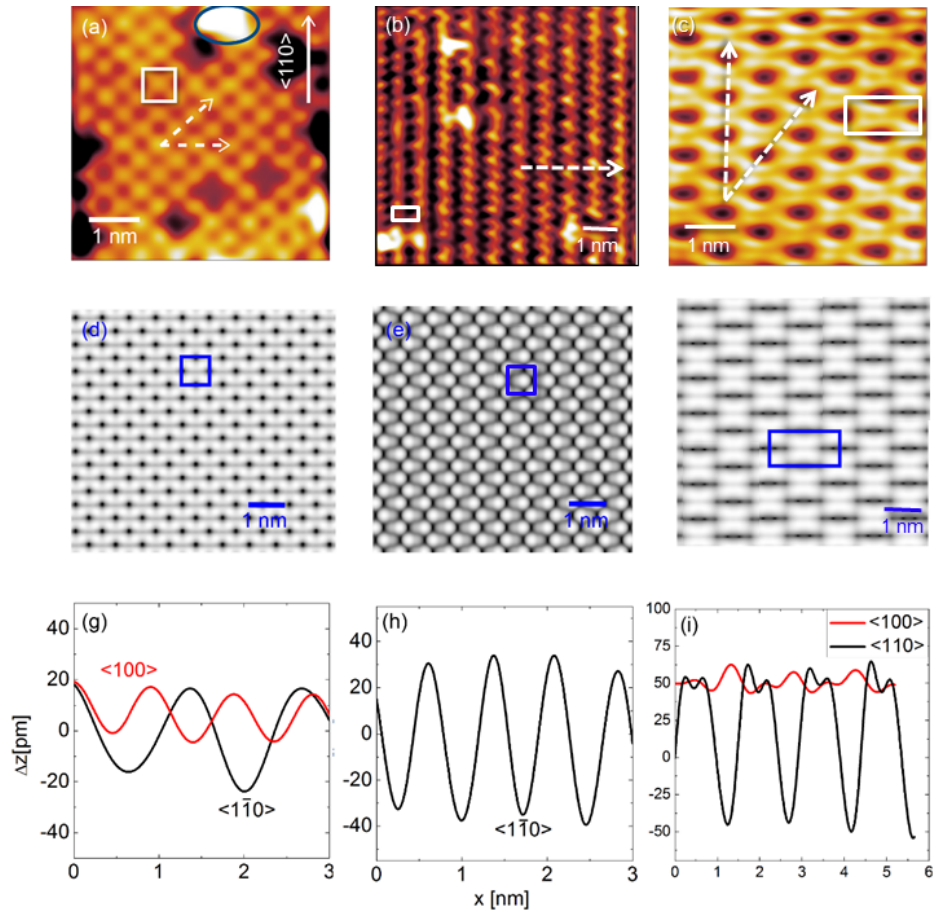

Figure 5. STM images of (a) c(2x2) (NiTi$_{0.97}$Sn$_{1.03}$), (b) (2x1), (2x2) (NiTiSn), and (c) c(2x4) (NiTi$_{1.01}$Sn$_{0.99}$) surfaces showing filled states at 77K using a W tip with a bias of -0.5V and a tunneling current of 30 pA, simulated STM images from DFT calculations for (d) c(2x2), (e) (2x2) and (f) c(2x4) surface reconstructions. Different simulated reconstructions are characterized by different Ti concentration on the surface. The smallest repeating unit cell in each of the cases is highlighted in the images. (g-i) Line scans of the corresponding surfaces to surface above, with dotted lines marking where depth profiles were taken, and solid squares highlighting surface unit cells. The c(2x4) and (2x1) surface also showed domains rotated 90° relative to those shown.

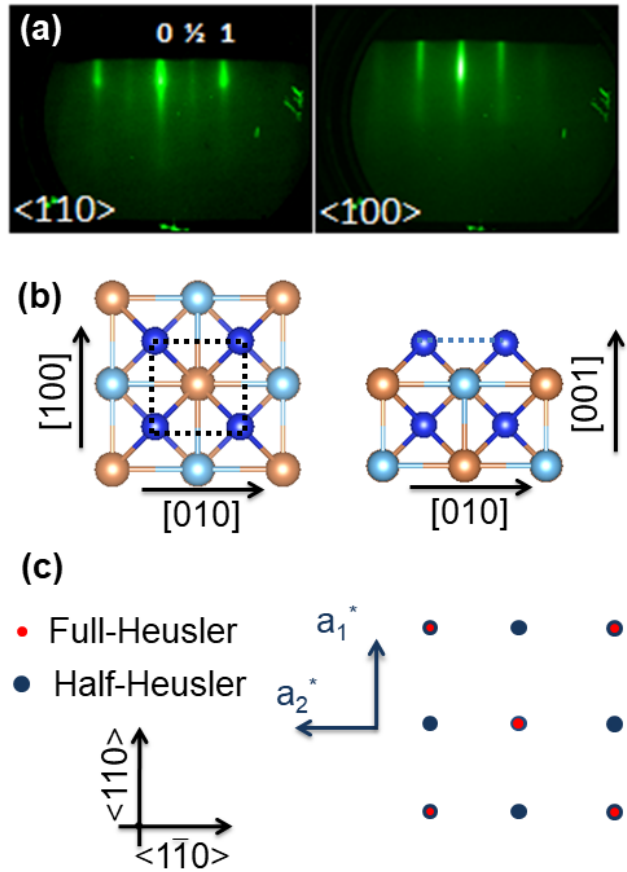

Figure 6. (a) RHEED of $Ni_2TiSn$ films (b) Model for a Ni-terminated $Ni_2TiSn$ surface (c) reciprocal space models of a Sn-Ti terminated half-Heusler and a Ni-terminated full-Heusler surface. $a_1^*$, $a_2^*$, and $b_1^*$, $b_2^*$, correspond to basis vectors for the half-Heusler surface unit cell and the full-Heusler surface unit cell, respectively

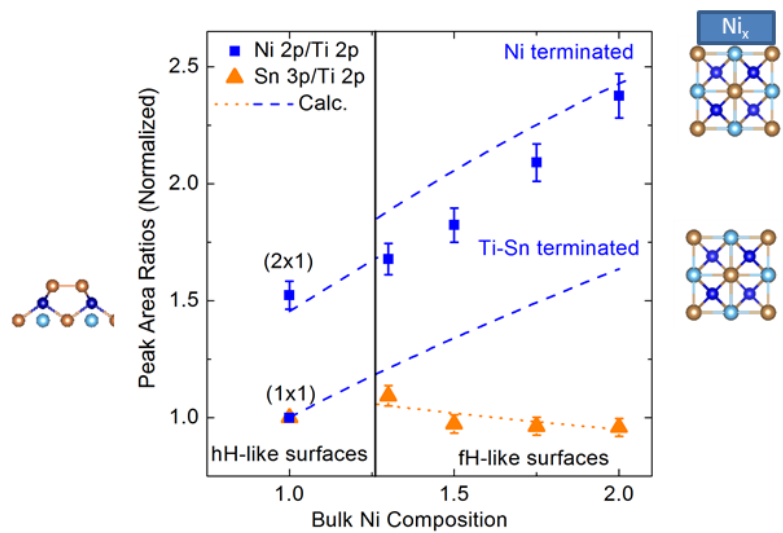

Figure 7. Integrated XPS peak area ratios of Ni 2p/Ti 2p and Sn 3p/Ti 2p as a function of Ni composition for $N_{1+x}TiSn$. Dotted lines represent expected values calculated assuming coverages shown, with 2 blue lines showing the predicted Ni/Ti ratio for two different assumptions of the surface termination. The top blue dashed line corresponds to $x/2$ monolayers of Ni termination (a Ni terminated surface) and the bottom blue line to 1 monolayer of Ti-Sn (a Ti-Sn terminated surface). Note that the orange dashed line corresponding to the predicted Sn/Ti ratio is for Ni termination but is nearly insensitive to the termination. The vertical black line corresponds approximately to the observed change in RHEED. The (2x1) normalized Sn/Ti ratio value is not shown (equal to 0.67). Schematics of surfaces are shown next to each model.

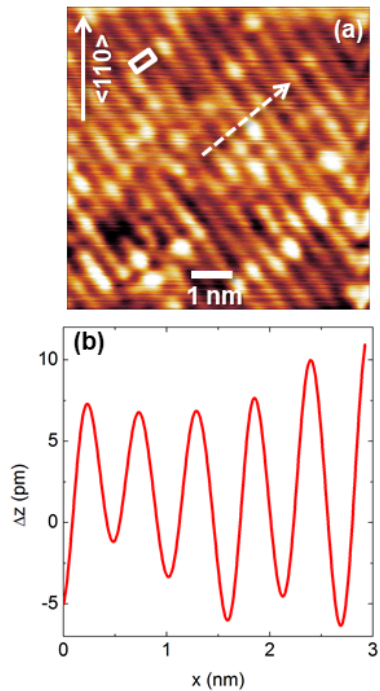

Figure 8. (a) STM of a full-Heusler (Ni$_2$TiSn surface showing filled states at 77K using a W tip with a bias of -0.5 V and a tunneling current of 30 pA and (b) step profiles taken along dotted line in 8(a). Other terraces on the surface showed similar rows rotated 90° relative to the one shown.

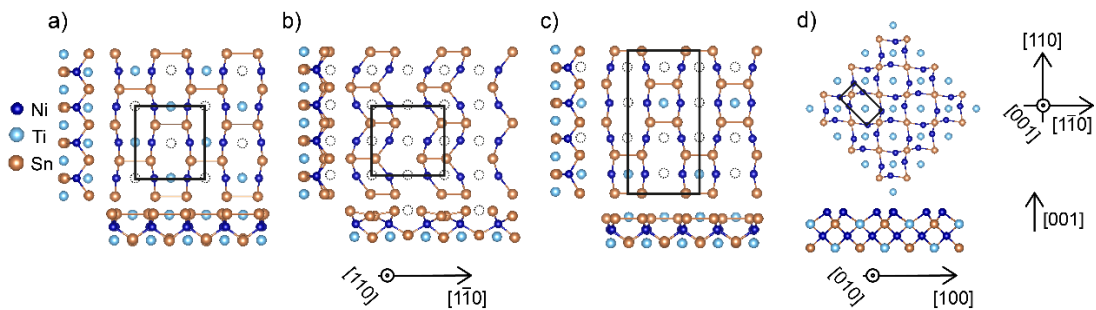

Figure 9. Ball-and-stick models for (a) c(2x2) (NiTi$_{0.97}$Sn$_{1.03}$), (b) (2x2) (NiTiSn), (c) c(2x4) (NiTi$_{1.01}$Sn$_{0.99}$ at low T) and (d) fH (Ni$_2$TiSn). Ti vacancies in the surface layer are highlighted by dotted circles in (a-c) (it should be noted that Sn atoms from lower layers are visible at these sites). A (2x1) model would be similar to the (2x2) model, with less dimer buckling in both directions, producing straighter rows.

## References

[1] J. Tobola, J. Pierre, S. Kaprzyk, R.V. Skolozdra, M.A. Kouacou, Crossover from semiconductor to magnetic metal in semi-Heusler phases as a function of valence electron concentration, J. Phys. Condens. Matter **10,** 1013 (1998) http://iopscience.iop.org/0953-8984/10/5/011

[2] R.A. deGroot, F.M. Mueller, P.G. van Engen, K.H.J. Buschow, New Class of materials: half-metallic ferromagnets, Phys. Rev. Lett. **50**, 2024 (1983) http://dx.doi.org/10.1103/PhysRevLett.50.2024.

[3] J. Winterlik, G. Fecher, A. Thomas, C. Felser, Superconductivity in palladium- based Heusler compounds, Phys. Rev. B **79,** 064508 (2009) http://dx.doi.org/ 10.1103/PhysRevB.79.064508.

[4] W. Al-Sawai, H. Lin, R. Markiewicz, L.A. Wray, Y. Xia, S.-Y. Xu, et al., Topological electronic structure in half-Heusler topological insulators, Phys. Rev. B **82**, 125208 (2010), http://dx.doi.org/10.1103/PhysRevB.82.125208.

[5] Palmstrøm CJ. Heusler Compounds and Spintronics. Prog. Crys. Growth Char. Mat.. **62**, 371 (2016) https://doi.org/10.1016/j.pcrysgrow.2016.04.020

[6] T. Graf, C. Felser, and S. S. P. Parkin, Simple rules for the understanding of Heusler Compounds, Prog. Solid State Chem. **39**, 1 (2011) https://doi.org/10.1016/j.progsolidstchem.2011.02.001

[7] T. M. Nakatani and A. Rajanikanth. Structure, magnetic property, and spin polarization of $Co_2FeAl_xSi_{1-x}$
Heusler alloys. J. Appl. Phys. **102**, 033916 (2007) http://dx.doi.org/10.1063/1.2767229

[8] Y. Nishino, H. Kato, M. Kato, and U. Mizutani. Effect of off-stoichiometry on the transport properties of the Heusler-type $Fe_2VAl$ compound. Phys. Rev. B **63**, 233303 (2001) https://doi.org/10.1103/PhysRevB.63.233303

[9] J.R. Arthur. Interaction of Ga and $As_2$ Molecular Beams with GaAs Surfaces. J. Appl. Phys. **39** (1968) 4032. https://doi.org/10.1063/1.1656901

[10] C.D. Theis, J.Yeh, D.G. Schlom, M.E. Hawley, G.W. Brown. Adsorption-controlled growth of $PbTiO_3$ by reactive molecular beam epitaxy. Thin Sol. Films **325,** 107 (1998) https://doi.org/10.1016/S0040-6090(98)00507-0

[11] J. K. Kawasaki, A. Sharan, L. I. M. Johansson, M. Hjort, R. Timm, B. Thiagarajan, B. D. Schultz, A. Mikkelsen, A. Janotti, and C. J. Palmstrøm, A simple electron counting model for half-Heusler surfaces, Science Advances **4,** eaar5832 (2018) https://doi.org/10.1126/sciadv.aar5832

[12] S.J. Patel, J. A. Logan, S.H. Harrington, B.D. Schultz, C.J. Palmstrøm, Surface reconstructions and transport of epitaxial PtLuSb (001) thin films grown by MBE, J. Crys. Growth **436**, 145 (2016) http://dx.doi.org/10.1016/j.jcrysgro.2015.12.003

[13] A.S. Bracker M.J. Yang, B.R. Bennett, J.C. Culbertson, W.J. Moore. Surface reconstruction phase diagrams for InAs, AlSb, and GaSb. J. Crys. Growth **220**, 384 (2000) http://dx.doi.org/10.1016/S0022-0248(00)00871-X

[14] A. Ohtake. Surface reconstructions on GaAs(001). Surface Sci. Rep. **63**, 295 (2008) doi:10.1016/j.surfrep.2008.03.001

[15] M. Hichour, D.Rached, R.Khenata, M.Rabah, M.Merabet, AliH.Reshak, S. BinOmran, R.Ahmed. Theoretical investigations of NiTiSn and CoVSn compounds. J. Phys. Chem. Solids **73**, 975 (2012) http://dx.doi.org/10.1016/j.jpcs.2012.03.014

[16] A.D. Rice, J.K. Kawasaki, N. Verma, D. J. Pennachio, B.D. Schultz, and C.J. Palmstrøm. Structural and Electronic Properties of Molecular Beam Epitaxially Grown $Ni_{1+x}TiSn$ Films. J. Crys. Growth **467C**, 71 (2017) ttp://dx.doi.org/10.1016/j.jcrysgro.2017.03.015

[17] J.K. Kawasaki, T. Neulinger, R. Timm, M. Hjort, A. a Zakharov, A. Mikkelsen, B.D. Schultz, and C.J. Palmstrøm. Epitaxial growth and surface studies of the half Heusler compound NiTiSn (001). J. Vac. Sci. Technol. B **31**, 04D106 (2013) http://dx.doi.org/ 10.1116/1.4807715.

[18] B.D. Schultz, H.H. Farrell, M.M.R. Evans, K. Ludge, and C.J. Palmstrøm. ErAs Inter-layers for Limiting Interfacial Reactions in Fe/GaAs (100) Heterostructures, J. Vac. Sci. Technol. B **20**, 1600 (2002) http://dx.doi.org/10.1116/1.1491994.

[19] S.H. Chen, C.B. Carter, and C.J. Palmstrøm. Lateral diffusion in Ni-GaAs couples investigated by transmission electron microscopy. J. Mater. Res. **3,** 1385 (1988) https://doi.org/10.1557/JMR.1988.1385

[20] S. Tanuma, C. J. Powell and D. R. Penn. Calculations of electron inelastic mean free paths. IX. Data for 41 elemental solids over the 50 eV to 30 keV range. Surf. Interface Anal **43** (2011) 689 https://doi.org/10.1002/sia.3522

[21] M. P. Seah, in Practical Surface Analysis, 2nd ed, edited by D. Briggs and M. P. Seah (Wiley, Chichester, 1990), Vol. 1.

[22] A. Seidl, A. Görling, P. Vogl, J. A. Majewski, and M. Levy. Generalized Kohn-Sham schemes and the band-gap problem, Phys. Rev. B **53**, 3764 (1996) https://doi.org/10.1103/PhysRevB.53.3764

[23] P. Hohenberg and W. Kohn. Inhomogeneous Electron Gas. Phys. Rev. **136**, B864 (1964) https://doi.org/10.1103/PhysRev.136.B864

[24] W. Kohn and L. J. Sham. Self-Consistent Equations Including Exchange and Correlation Effects. Phys. Rev **140**, A1133 (1965) https://doi.org/10.1103/PhysRev.140.A1133

[25] J. P. Perdew, K. Burke, and M. Ernzerhof. Generalized Gradient Approximation Made Simple. Phys. Rev Lett. **77** (1996) 3865 https://doi.org/10.1103/PhysRevLett.77.3865

https://link.aps.org/doi/10.1103/PhysRevLett.77.3865.

[26] G. Kresse and J. Furthmüller. Efficient iterative schemes for ab initio total-energy calculations using a plane-wave basis set. Phys. Rev. B **54**, 11169 (1996) https://doi.org/10.1103/PhysRevB.54.11169

[27] G. Kresse and J. Furthmüller. Efficiency of ab-initio total energy calculations for metals and semiconductors using a plane-wave basis set. Comp. Mat. Sci. **6,** 15 (1996) https://doi.org/10.1016/0927-0256(96)00008-0

[28] P. E. Blöchl. Projector augmented-wave method. Phys. Rev. B **50**, 17953 (1994) https://doi.org/10.1103/PhysRevB.50.17953

[29] G. Kresse and D. Joubert. From ultrasoft pseudopotentials to the projector augmented-wave method. Phy. Rev. B B **59**, 1758 (1999) https://doi.org/10.1103/PhysRevB.59.1758

[30] M. Hichour, D.Rached, R.Khenata, M.Rabah, M.Merabet, AliH.Reshak, S. BinOmran, R.Ahmed, J. Phys. Chem. Solids **73** (2012) 975 http://dx.doi.org/10.1016/j.jpcs.2012.03.014

[31] Danny E.P. Vanpoucke and Geert Brocks. Formation of Pt-induced Ge atomic nanowires on Pt/Ge(001): A density functional theory study. Phys. Rev. B **77** (2008) 241308(R)

URL https://journals.aps.org/prb/abstract/10.1103/PhysRevB.77.241308

[32] D. I. Lubyshev, M. Micovic, D. L. Miller, I. Chizhov and R. F. Willis. Molecular beam epitaxial growth of InAs on a (311)A corrugated surface: Growth mechanism and morphology. J. Vac. Sci. Technol. B **16** (1998) 1339 doi: 10.1116/1.590071

[33] V.V. Romaka, P. Rogl, L. Romaka , Yu.Stadnyk, N. Melnychenko, A. Grytsiv, M. Falmbigl, and N. Skryabina. Phase equilibria, formation, crystal and electronic structure of ternary compounds in Ti–Ni–Sn and Ti–Ni–Sb ternary systems, J. Solid State Chem. **197**, 103 (2013) http://dx.doi.org/10.1016/j.jssc.2012.08.023.

[34] H. Kandpal, C.Felser, and R.Seshadri,J. Phys.D39,776 (2006) http://dx.doi.org/10.1088/0022-3727/39/5/S02

[35] J.A. Logan, T.L. Brown-Heft, S.D. Harrington, N. Wilson, A.D. Rice, and C.J. Palmstrøm. Growth, Structural, and Magnetic Properties of single-crystal full-Heusler $Co_2TiGe$ Thin Films, J. App. Phys. **121**, 213903 (2017) https://doi.org/10.1063/1.4984311

[36] Y. Sakuraba, T. Iwase, K. Saito, S. Mitani, and K. Takanashi. Enhancement of spin-asymmetry by $L2_1$-ordering in $Co_2MnSi/Cr/Co_2MnSi$ current-perpendicular-to-plane magnetoresistance devices. Appl. Phys. Lett. **94**, 012511 (2009) http://dx.doi.org/10.1063/1.3068492

[37] A. Neggache, T. Hauet, F. Bertran, P. Le Fèvre, S. Petit-Watelot, T. Devolder, P. Ohresser, P. Boulet, C. Mewes, S. Maat, J. R. Childress, and S. Andrieu. Testing epitaxial $Co_{1.5}Fe_{1.5}$ Ge(001) electrodes in MgO-based magnetic tunnel junctions. Appl. Phys. Lett. **104**, 252412 (2014) http://dx.doi.org/10.1063/1.4885354